\documentclass[psfig]{kluwer}    
\usepackage{graphicx}
\newdisplay{guess}{Conjecture}
\begin{document}                                                                                   
\begin{article}
\begin{opening}         
\title{Giant Meterwave Radio 
Telescope observations of an M2.8 flare: insights into the initiation of a flare-coronal mass ejection event} 
\author{Prasad \surname{Subramanian}\thanks{Inter-University Centre for Astronomy and Astrophysics, P.O Bag 4, Ganeshkhind, Pune - 411007, India. e-mail: psubrama@iucaa.ernet.in}, S. \surname{Ananthakrishnan}\thanks{National Centre for Radio Astrophysics, TIFR, P. O. Bag 3, Ganeshkhind, Pune - 41007, India}, P. \surname{Janardhan}\thanks{Physical Research Laboratory, Astronomy \& Astrophysics Division, Navrangpura, 
Ahmedabad - 380009, India}}  
\runningauthor{Subramanian et al.}
\runningtitle{Initiation of flare-CME event}
\institute{}
\author{M. R. Kundu, 
S. M.White, 
V. I. Garaimov \thanks{Dept. of Astronomy, University of Maryland, College Park, MD 20742, USA}}
\institute{}


\begin{abstract}
We present the first observations of a solar flare with the GMRT. 
An M2.8 flare observed at 1060 MHz with
the GMRT on Nov 17 2001 was associated with a prominence eruption observed at 17 GHz by the 
Nobeyama radioheliograph and the initiation of a fast partial halo CME observed with
the LASCO C2 coronograph. Towards the start of the eruption, we find evidence
for reconnection above the prominence.
Subsequently, we find evidence for rapid growth of a
vertical current sheet below the erupting arcade, which is accompanied by the flare and prominence eruption.  
\end{abstract}

\end{opening}           

\section{Introduction} 

We have recently started observing the Sun with the Giant Meterwave Radio Telescope (GMRT) near
Pune, India. We report here the observation of an interesting flare-coronal mass ejection (CME) 
event on Nov 17 2001.
The event was analyzed using data from the GMRT, the Nobeyama Radioheliograph (NoRH) and the Hiraiso 
Spectrograph (HiRAS) in Japan, the Radio Solar Telescope Network (RSTN) in Australia, the Mount Wilson
observatory in the USA and the Michelson Doppler Imager (MDI) and 
Large Angle Spectroscopic Coronograph (LASCO) onboard the Solar and 
Heliospheric Observatory (SOHO). 

We give a brief description of the GMRT and the solar observing procedure in \S 2, followed by  
a multiwavelength overview of the flare-CME event in \S 3.  We interpret the data in \S 4 and 
draw conclusions in \S 5.

\section{The Giant Metrewave Radio Telescope (GMRT)}
The GMRT is located about 80km north of the city of Pune in Maharashtra, western India 
(latitude 19$^{o}\,06^{'}$N, longitude 74$^{o}\,03^{'}$E, altitude 650 meters).
It consists of thirty 45m diameter antennas spread over 25 km. Half of these are in a
compact, randomly distributed array of about 1 km and the rest are spread out in an approximate `Y'
configuration. The shortest baseline is 100m and 
the longest is 26 km. GMRT currently operates in the frequency bands
around 150, 235, 325, 610 and 1000--1450 MHz. It is among the most 
sensitive telescopes in the world in these bands and also provides
seconds of arc angular resolution (Ananthakrishnan \& Rao, 
2002). Further details about the GMRT can be found at the URL
http://www.gmrt.ncra.tifr.res.in
\subsection{Observing the Sun with the GMRT}
These observations represent the first attempts at imaging the Sun with the
GMRT after most of the antennas had started functioning. Our observing
frequencies ranged from 1060 to 1076 MHz.
There were 28 antennas working through
the duration of the observations presented in this paper. We did not track the Sun
continuously during these observations. Instead, we repositioned the antennas at the
Sun center every 30 minutes. Within each 30 minute interval, the
phase center was allowed to drift. The data were then corrected for solar rotation during
subsequent data analysis. We typically observed the
Sun for around 20 minutes and the phase calibrator (3C298) for 6-7 minutes in each
30 minute cycle. Solar attenuators of 30 dB were inserted while observing the Sun
and they were removed while recording data on the calibrator.
Since the GMRT antennas are not yet equipped with extra high noise calibrators (which can
generate noise corresponding to $\sim$ 1 sfu) we cannot place much confidence in the
absolute values of flux observed at this time. We therefore normalized the fluxes we observed to
the peak flux observed by the Nobeyama Radio Polarimeter (NoRP) at 
1 GHz (http://solar.nro.nao.ac.jp/norp/html/daily/).
Since the automatic gain control (AGC) circuits were
on throughout the duration of the observations, the flux at the final peak of the flare
is somewhat supressed relative to the preceding smaller peaks (see figure 8).
We used an integration time of 16 seconds; this is therefore the minimum time
cadence with which we can obtain snapshot images for this observation. 
In principle, it is possible to go down to an integration time as low as 2 seconds  
but no lower, because the time constant of the AGC circuits is of the order of 1 second. 
The data were processed using the standard NRAO/AIPS software. 
Snapshot images, with a dynamic range of $\simeq$10, of the flaring region were obtained 
every 16 seconds using a restoring beam size of 34$^{''} \times$ 24$^{''}$ in right ascension and 
declination respectively. 

\section{An Overview of the event}

A long duration M2.8 flare was observed on 17 Nov. 2001 at 1060 MHz with the GMRT. The 
X-ray flare which originated in NOAA active region 9704 (S18E41), 
started at 04:49 UT, showed a smooth rise,  
%
%
\protect\begin{figure}[ht]
\vspace{19pc}
\caption{ Soft X-ray lightcurves from the GOES satellites. Upper curve: 1 -- 8 \AA. Lower curve: 0.5 -- 4 \AA.
The X-ray flare started at 04:49 UT, showed ~a smooth rise and peaked at 05:25 UT 
and exhibited a long decay period before ending at 06:11 UT.}
\label{goes}
\includegraphics{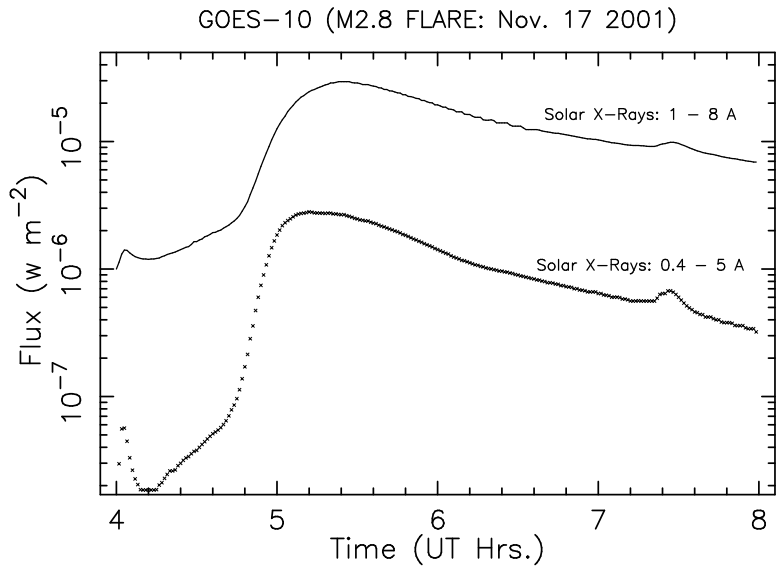}
\end{figure} 
%

\noindent peaked at 05:25 UT 
and exhibited a long decay period before ending at 06:11 UT.  The X-ray fluxes 
observed by GOES 10 in the 1-8 $\AA$  and 0.5-4 $\AA$ bands are shown in Figure 1. 
 Figure 2 shows the MDI longitudinal magnetogram of AR 9704 (left panel) and the Mount Wilson 
%
\protect\begin{figure}[ht]
\vspace{17pc}
\caption{{\em Left Panel}: The MDI longitudinal magnetogram of AR 9704. The magnetic complexity of AR 9704 is
clearly evident. {\em Right Panel}: The Mount 
Wilson sunspot drawing for this region. The Mount Wilson sunspot drawing 
on Nov 17 2003 assigns a $\beta\,\gamma\,\delta$ category to this
active region. }
\label{mdi}
\includegraphics{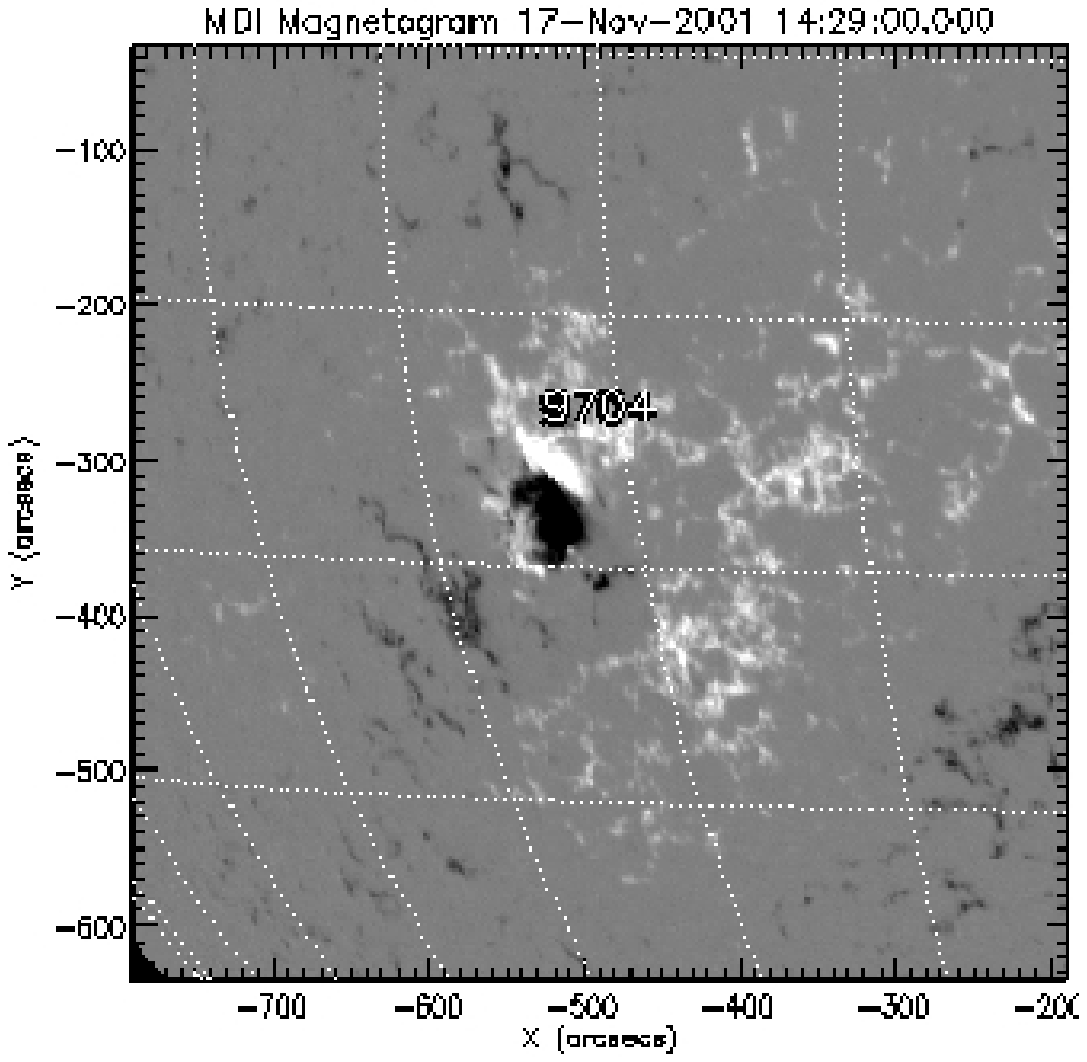}
\includegraphics{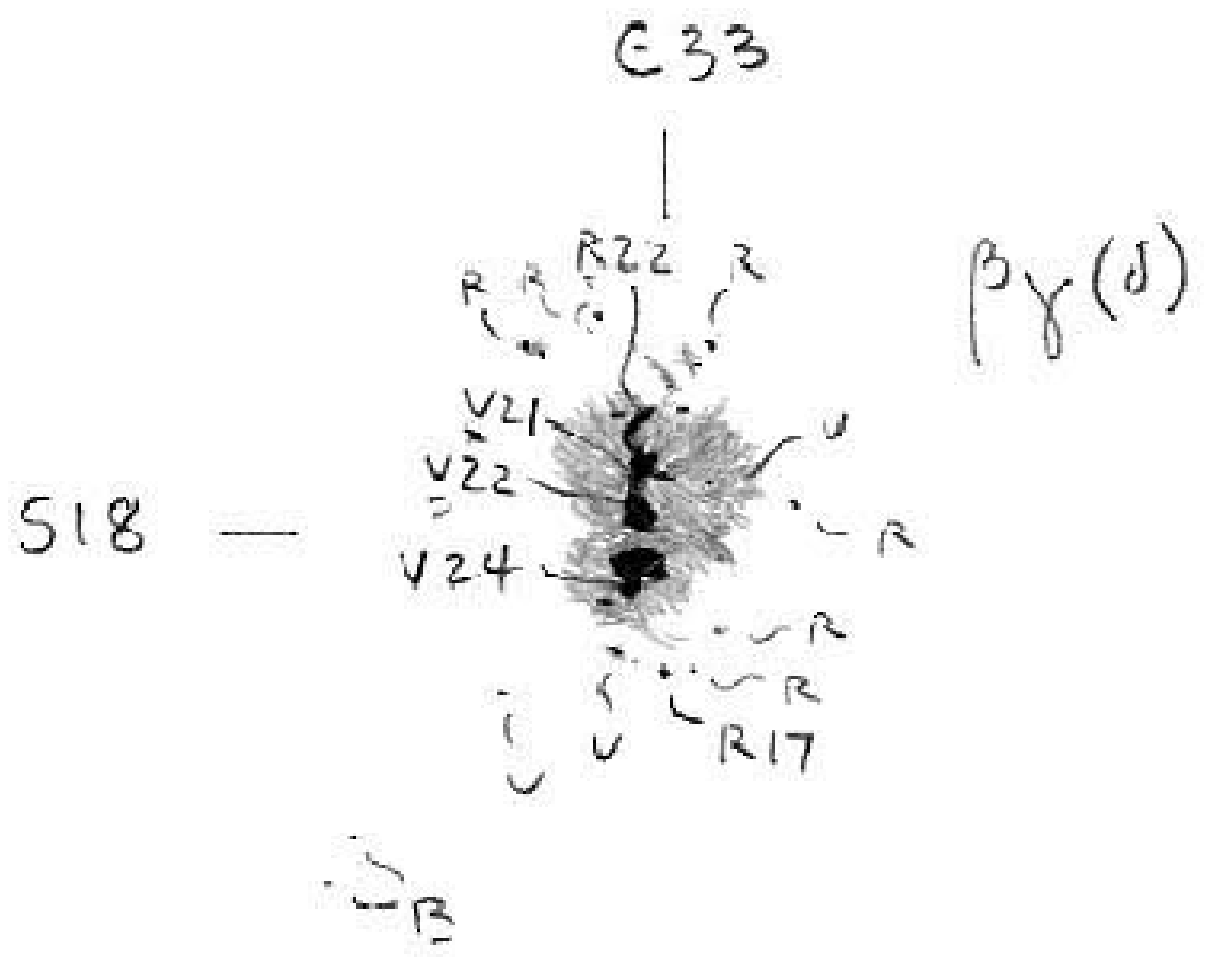}
\end{figure} 
%
%
sunspot drawing for this region (right panel). An examination of the longitudinal magnetogram shows that 
it is clearly more complex than a simple bipole. The central black polarity spot is sandwiched by 
white polarities to its northwest and southeast, suggesting the existence of 
at least two sets of loop systems. This active region was classified as a $\beta$ configuration by 
the Solar Geophysical Data Reports 
on Nov 17 2001, but its
complexity was upgraded to $\beta\,\gamma\,\delta$ on Nov 19 2003, when it had rotated closer to disk
center. The Mount Wilson sunspot drawing on Nov 17 2003 
assigns a $\beta\,\gamma\,\delta$ category to this
active region. Notwithstanding the disparity between the classifications of AR 9704 on Nov 17 2003, 
the magnetogram
and the sunspot drawing provide sufficient evidence for the magnetic complexity of AR 9704. The 1060 MHz 
GMRT images, described later, (see figure 9) show evidence for two sets of loop systems in this region, 
providing additional evidence for its complexity.

The flare was accompanied by a 
fast halo ($\sim$ 1500 km/s) CME that first appeared in the C2 field of view of LASCO 
at around 05:30 UT (http://cdaw.gsfc.nasa.gov/CME$_{-}$list). 
Figure 3 shows the lightcurve of the 17 and 34 GHz emission from the NoRH and a representative 17 GHz image 
%
\protect\begin{figure}[h]
\vspace{29pc}
\caption{{\em Upper panel}: The upper curve is the 17 GHz lightcurve and the lower curve is the
34 GHz lightcurve from the NoRH. The vertical axis gives the flux in SFU. 
Both the lightcurves show a gradual rise and decay, implying that
the radiation at these frequencies is thermal in nature. This radiation arises primarily from the
active region situated at S18E40, that is evident in the lower panel.
{\em Lower panel}: Image of the Sun at 17 GHz at 04:55 on Nov 17 2001 from 
the NoRH showing the erupting
prominence on the east limb. The prominence, which was situated towards the northeast of the
thermal active region emission evident at S18E40, erupted at $\sim$ 04:48 UT. 
A movie of this event is available at 
http://solar.nro.nao.ac.jp/norh/html/10min/}
\label{filament}
\includegraphics{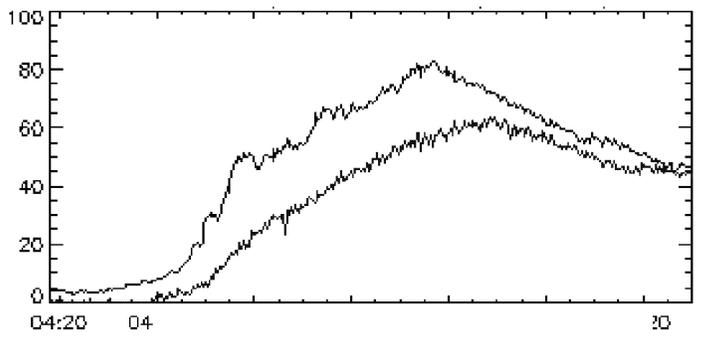}
\includegraphics{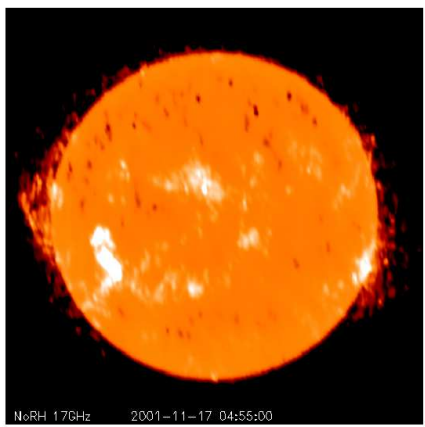}
\end{figure} 
%
%
from the NoRH that shows the erupting prominence. The bright 
on-disk emission at 17 GHz
evident in Fig.\ 3 is centered at around S18E40. The 17 GHz emission
shows a characteristically gradual rise and decay, indicating that it is
thermal in nature at these frequencies.  On the other hand, the impulsive nature of the 
200, 410, 500 and 1060 MHz lightcurves (see figures 6--8) is evidence of their nonthermal nature. 
The 10 minute cadence movie of the 17 GHz data for this event available on 
the Nobeyama webpage (see figure 3) suggests that the prominence started lifting off sometime between
04:45 UT and 04:55 UT. A detailed analysis of the prominence eruption associated with this event  
by Kundu et al. (2003) establishes that the rapid phase of the prominence
liftoff started at around 04:48 UT. 

Figure 4 shows dynamic spectrum of the event from the Hiraiso spectrograph.  The spectrum shows
%
%
\protect\begin{figure}[ht]
\vspace{19pc}
\caption{Dynamic spectra from the Hiraiso spectrograph. 
The start of the continuum emission at $\sim$ 200 MHz at $\sim$ 04:35 UT
is clearly evident. The lower envelope of the drifting continuum represents the upper 
end of a feature that is drifting upwards. The height-time profile of this lower envelope 
is shown in figure 5. 
}
\label{dynspec}
\includegraphics{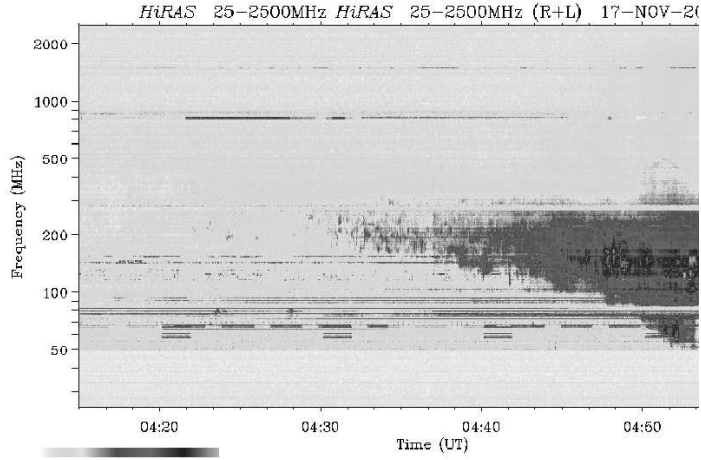}
\end{figure} 
%
%
drifting continuum emission starting at $\sim$ 04:35 UT at $\sim$ 200 MHz and drifting downwards in frequency. 
The lower envelope of this drifting continuum emission represents the 
upper end of a feature that is drifting upward in the solar corona. The HiRAS dynamic spectrum did 
not have data below 50 MHz for this
event. We were therefore able to follow the lower envelope of the drifting continuum only until
$\sim$ 04:50 UT. An approximate curve was fitted to the lower envelope of the drifting continuum. 
Using the formulae given by Aschwanden \& Benz (1995), the frequency of emission was related to
its height in the solar atmosphere assuming that the observed radiation is 
emitted at the fundamental plasma frequency. 
Figure 5 shows the results of these computations. It shows the upward
%
\protect\begin{figure}[ht]
\vspace{22pc}
\caption{{\em Left panel}: The height-time profile of 
the lower envelope of the drifting continuum evident in the dynamic spectrum of figure 4. It 
represents the upper end
of a vertical current sheet that forms beneath the erupting prominence.
{\em Right panel}: The velocity-time profile of the same feature. The abrupt increase in velocity at around 
04:48 UT can be inferred from the increased curvature around this time 
of the lower envelope of the drifting continuum (figure 4).
}
\label{height-time}
\includegraphics{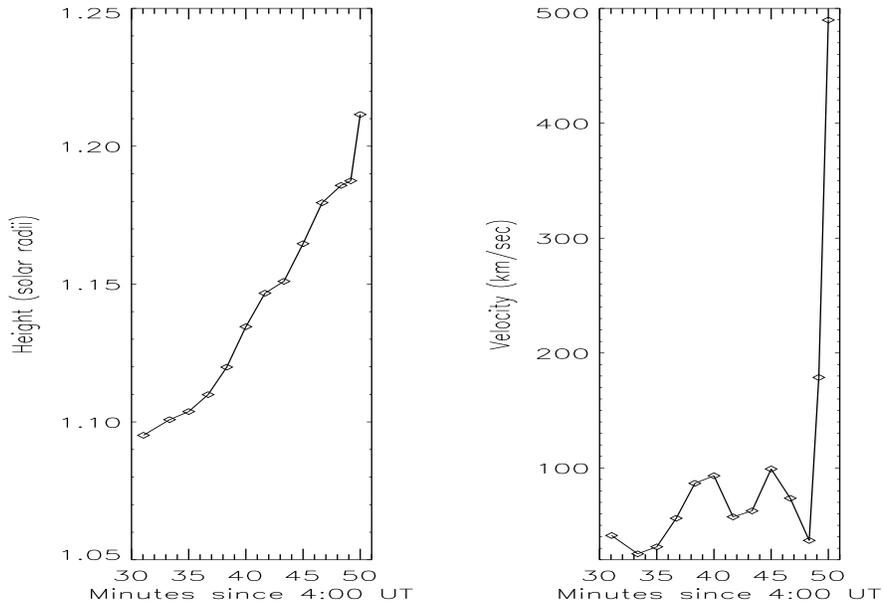}
\end{figure} 
%
%
motion of the feature represented by the lower envelope of the drifting continuum (evident in figure 4) 
through the 
solar corona.  

In contrast to the slow rise and fall in the 17 and 34 GHz light curves, the light curves at 
200, 410, 500 and 1060 MHz show an impulsive nature.  In fact, the 200 MHz emission is the 
first sign of non thermal activity associated with this event.  Figure 6 shows 
fixed frequency lightcurves at 200 MHz (lower panel) and 500 MHz (upper panel) from the Hiraiso 
spectrograph while Figure 7 shows the fixed frequency light curve at 410 MHz from the 
Radio Solar Telescope Network (RSTN) telescope at Learmonth, Australia. The emission at 200 Mhz starts 
rising at around 04:22 UT while that at 500 MHz starts rising at around 04:50 UT.  
%
%
\protect\begin{figure}[ht]
\vspace{30pc}
\caption{Fixed frequency lightcurves at 200 and 500 MHz from the Hiraiso spectrograph. The emission at
200 Mhz starts rising at around 04:22 UT while that at 500 MHz starts rising at around 04:50 UT.}
\label{hiras}
\includegraphics{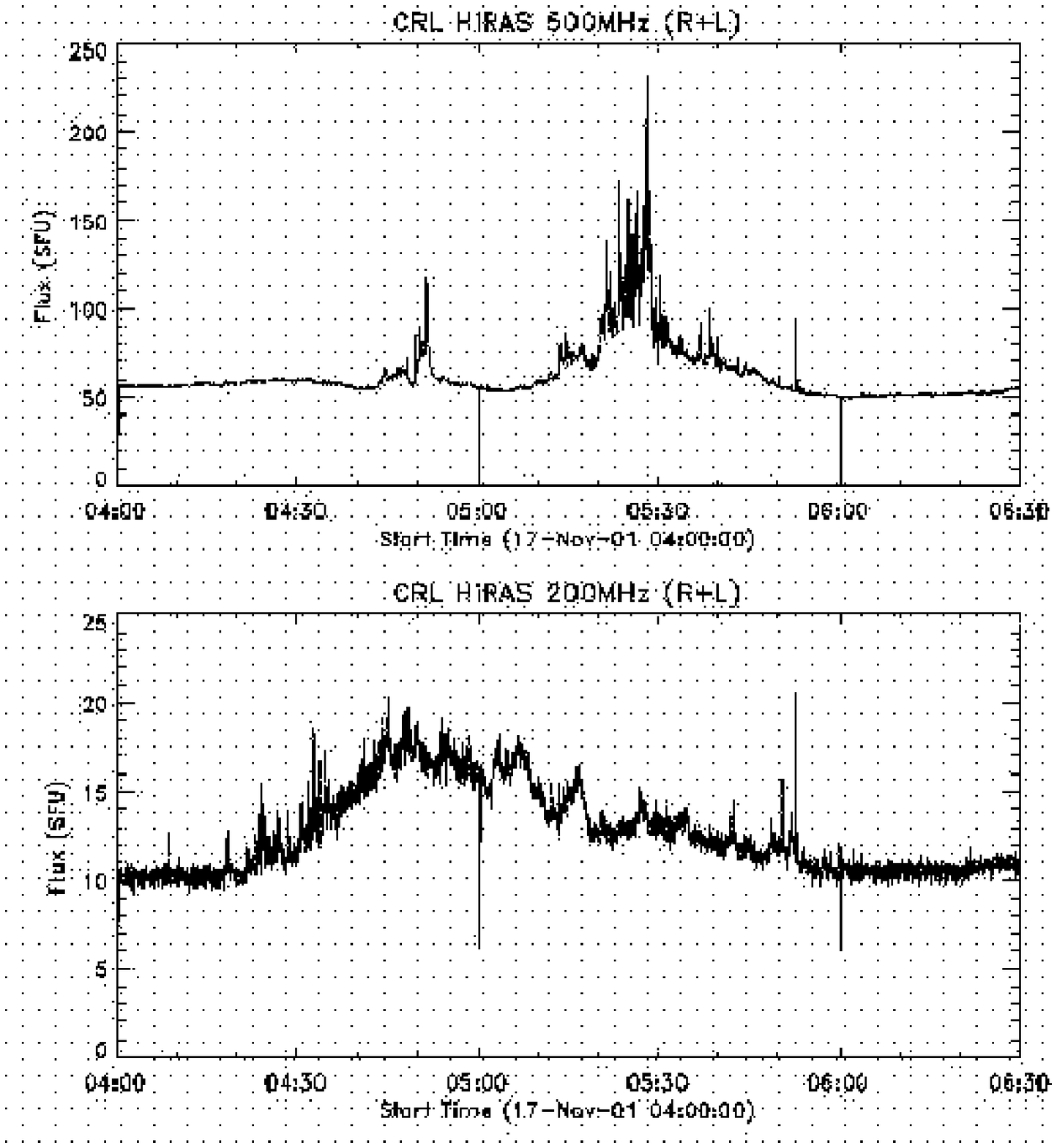}
\end{figure} 
%
%
\protect\begin{figure}[ht]
\vspace{23pc}
\caption{Fixed frequency light curve at 410 MHz from the 
Radio Solar Telescope Network (RSTN) telescope at Learmonth, Australia. The vertical axis is 
calibrated in SFUs above the baseline.}
\label{learmonth}
\includegraphics{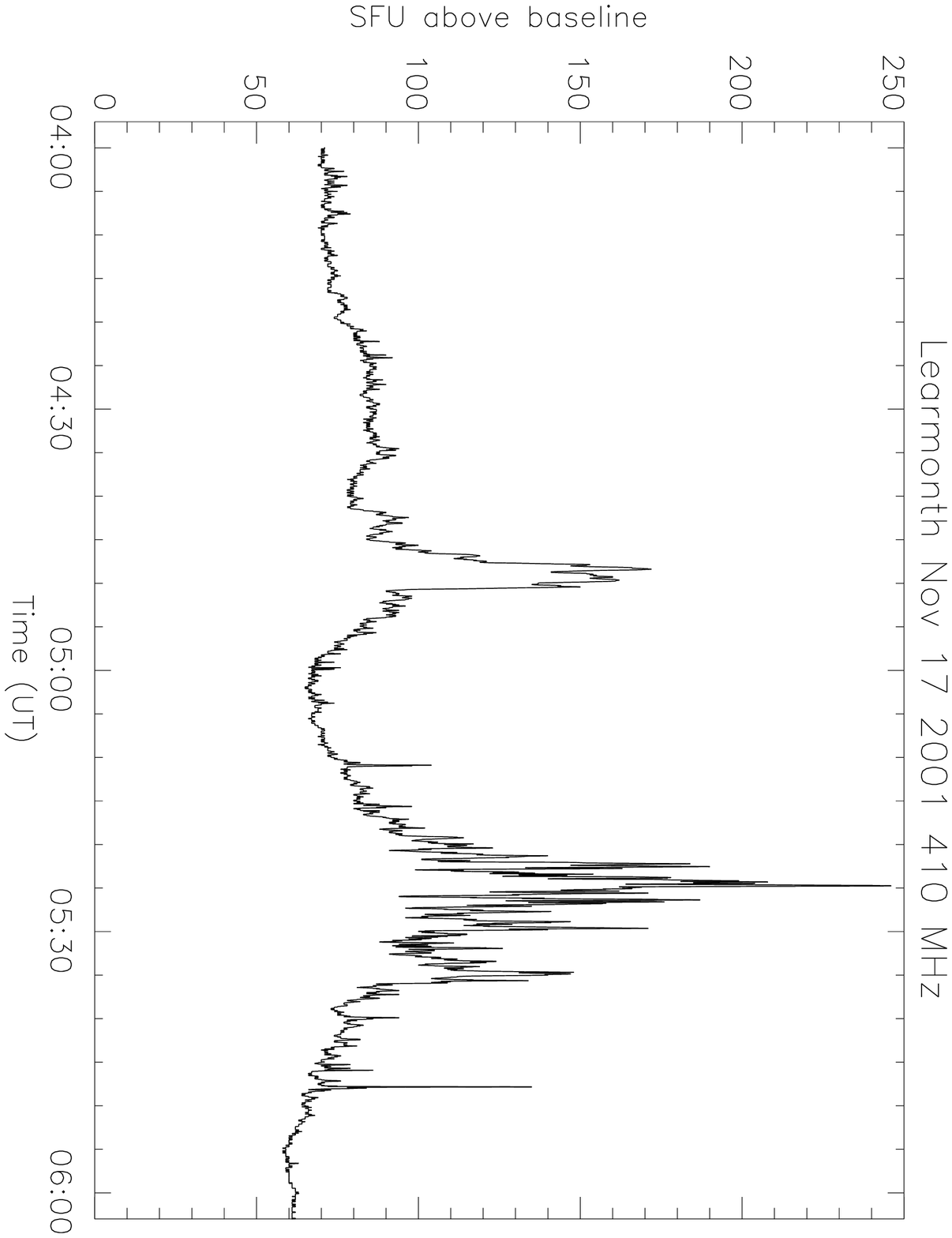}
\end{figure} 
%
%
Figure 8 shows the 1060 MHz light curve derived from the GMRT observations.
%
%
\protect\begin{figure}[ht]
\vspace{20pc}
\caption{The 1060 MHz lightcurve from GMRT observations. The three peaks are clearly evident. The
first onset of nonthermal emission at this frequency occurs at 04:48 UT. The peak of this lightcurve
is normalized with respect to the peak of the 1 GHz lightcurve from the 
NoRP (http://solar.nro.nao.ac.jp/norp/html/daily/). Figure 6 shows 16 s snapshot
images near the three peaks of this lightcurve.}
\label{lightcurve}
\includegraphics{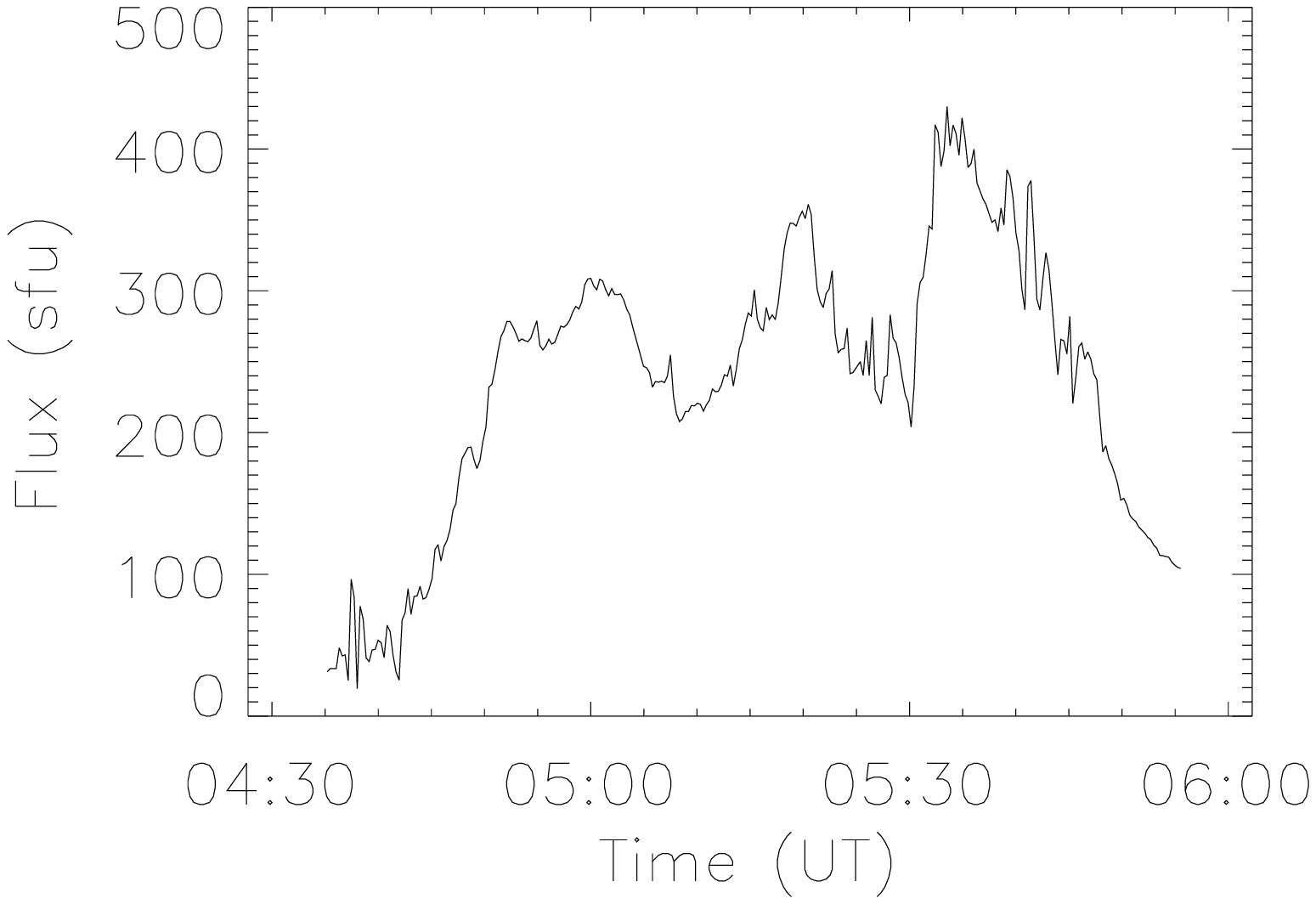}
\end{figure} 
%
%
This emission has three distinct peaks. The largest peak of the GMRT 1060 MHz lightcurve is
normalized to the largest peak of the 1 GHz lightcurve from the NoRP 
(http://solar.nro.nao.ac.jp/norp/html/daily/). The general shape of the GMRT 1060 MHz lightcurve is very
similar to the NoRP 1 GHz lightcurve. However, the third and largest peak
in the GMRT 1060 MHz data is suppressed relative to the first two peaks, since the AGC
circuits were on during the observation. As a result, the ratios peak1/peak3 and peak2/peak3 from the
GMRT 1060 MHz lightcurve are around
50\% larger than the corresponding ratios obtained from the NoRP 1 GHz lightcurve.
Figure 9 shows 16 s snapshots of the 1060 MHz emission observed with the
%
%
\protect\begin{figure}[ht]
\vspace{22pc}
\caption{Contours of 16 s snapshots of the GMRT 1060 MHz emission. The center of these images is
at S15E46 and the field of view is 500$^{''} \times$500$^{''}$. The restoring beam is 
34$^{''} \times$ 24$^{''}$.  The contour levels are 0.1, 0.2, 0.3, 0.4, 0.5, 0.6, 0.7, 0.8, 0.9 and 1.0 times
the flux at the largest peak, which occurs at 05:33 UT. We show four sets of 16 s snapshots starting at
04:52:23 UT, 05:20:28 UT and 05:33:32 UT. These correspond to the three peaks of the 1060 MHz 
lightcurve (figure 8).
The emission at the peaks of 04:52 UT and 05:20 UT 
are from a NS oriented loop system, whereas the emission at the peak at 05:33 UT is
from a NE-SW oriented loop. The brightness temperature of the emission at 
each of these peaks is $\lesssim 10^{8}$K.}
\label{snapshots}
\includegraphics{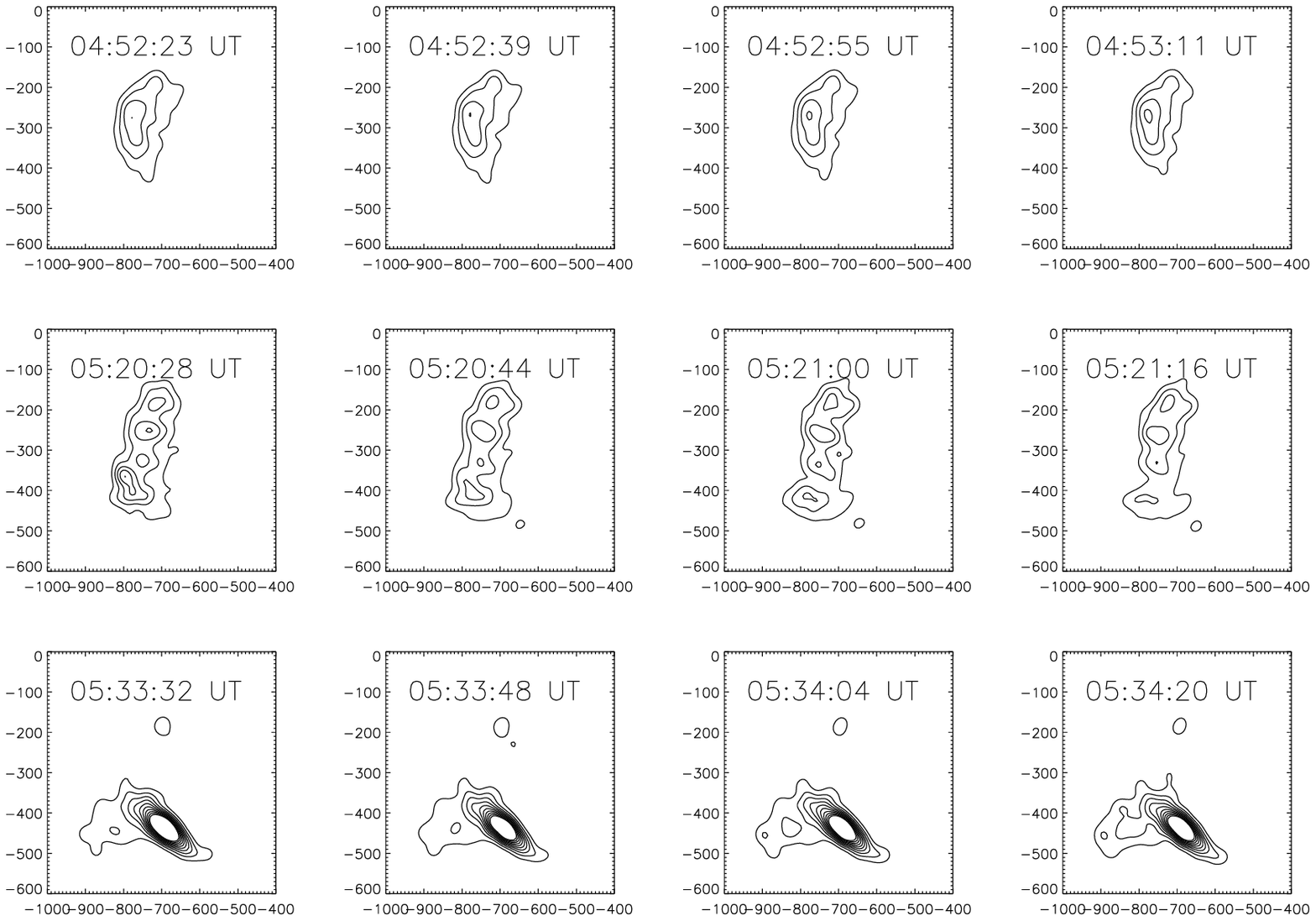}
\end{figure} 
%
%
GMRT. The images are centered at S15E46, with a field of view of 500$^{"} \times$ 500$^{"}$.
The first peak of the 1060 MHz GMRT lightcurve near 04:52 UT is associated with the brightening of 
a banana-shaped NS oriented structure near the center of the field of view of the 
images. The second peak near 05:18 UT is also associated with the brightening of 
the same structure. The third and strongest peak, at around 05:32 UT, corresponds to a brightening of a 
NE-SW oriented source to the south of the 'banana'. The brightness temperature of
the radiation at each of these three peaks is $\lesssim 10^{8}$K, indicating its
nonthermal nature. Figure 10 shows 16 s snapshot images near the rising edge of the first peak. The 
images clearly show that the first peak of the 1060 MHz emission occured at around 04:48 UT.

The 1060 MHz GMRT images reveal the existence of at least two sets of loop systems (N-S and NE-SW). 
Together with the magnetograms and sunspot drawing (see figure 2), this provides concrete  
evidence for the complexity of the magnetic fields  in the region from which the flare and CME were 
initiated.

%
\protect\begin{figure}[ht]
\vspace{21pc}
\caption{Snapshots at 16 second intervals near the rising edge of the first peak in the GMRT light curve. 
The contour levels and 
restoring beam are the same as those in figure 9. It is
clearly evident that the first peak of the 1060 MHz emission occurs between 04:47:32 and 04:47:48 UT.}
\label{snapshots2}
\includegraphics{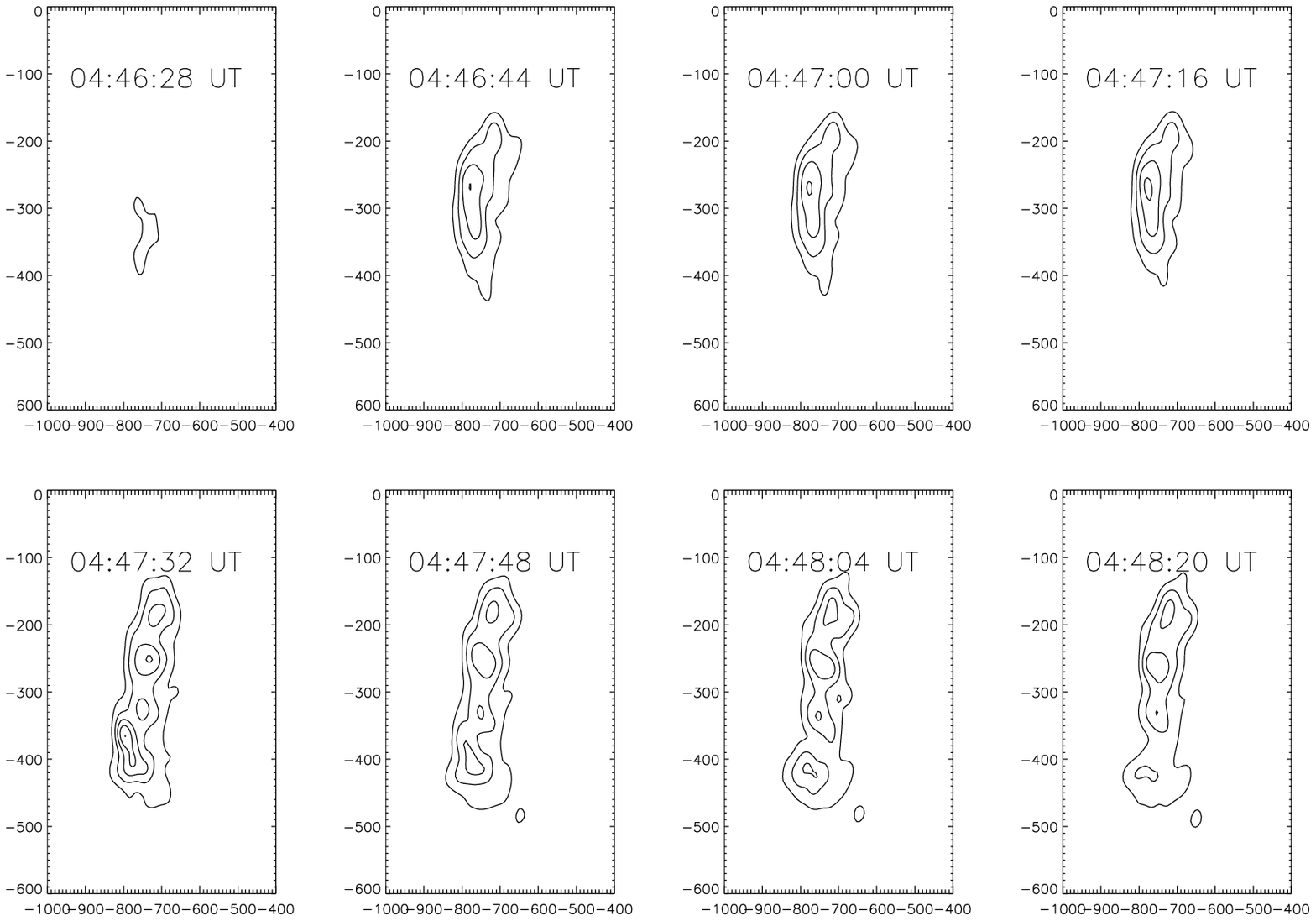}
\end{figure} 
%
%

\section{Initiation of the flare-CME event: analysis}

\begin{table*}
	\begin{tabular}{clccc}
	\hline
 ITEM & DESCRIPTION  &  R$_{\odot}$ & R$_{\odot}$ & START \\ 
  &  &  (s=1)& (s=2)& (UT) \\ \hline

  A & 200 MHz\endnote{Peak in fixed frequency lightcurve from Hiraiso, figure 6} & 1.08 &1.14 & 04:22   \\ 
  B & Lower envelope of & & & \\
    & drifting continuum\endnote{Rapid increase in velocity of lower envelope of drifting \\ 
\hspace*{5mm} continuum (Figs.\ 4 \& 5 )} &1.186 &1.33& 04:48 \\
  C & 410 MHz \endnote{Peak in fixed frequency lightcurve from Learmonth, figure 7} &1.044 &1.08&04:48 \\
  D & 1060 MHz N-S source (1)\endnote{GMRT 1060 MHz data; figures 8 and 10} & 1.02 & 1.04& 04:48 \\ 
  E & Prominence eruption\endnote{NoRH 17 GHz data; figure 3 and Kundu et al. (2003)}    & & & $\sim$ 04:48 \\ 
  F & X-ray flare\endnote{GOES data; figure 1, lower curve, and Masuda et al. (1994)}  &1.02 &1.02& 04:49  \\ 
  G & 500 MHz (2)\endnote{Peak in fixed frequency lightcurve from Hiraiso;  figure 6} &1.037&1.067 & 04:50 \\  
  H & 1060 MHz N-S source (2)\endnote{GMRT 1060 MHz data; figures 8 and 9} &1.02&1.04& 05:13  \\   
  I & 500 MHz (2)\endnote{Fixed frequency lightcurve from Hiraiso; figure 6} &1.04&1.06 & 05:13 \\  
  J & 1060 MHz NE-SW source\endnote{GMRT 1060 MHz data;  figures 8 and 9} &1.02&1.04 & 05:30 \\ 
  K & CME\endnote{LASCO C2 data}  & 5 &5& 05:30  \\ \hline

	\end{tabular}
\caption[]{Sequence of events in CME-flare initiation}

\theendnotes
\end{table*}

Table 1 shows the time sequence of events leading to, and after the CME-flare initiation, 
with emphasis on the nonthermal emission. The third and fourth columns
show the estimated height in $R_{\odot}$ from which the corresponding phenomenon is
believed to be emanating. We employ the model of Aschwanden \& Benz (1995) in calculating these heights. 
The third column shows the height assuming that the radio emission
emanates at the fundamental plasma frequency (s=1) while the height in the fourth column is 
calculated assuming that the radio emission emanates at the first harmonic of the plasma frequency (s=2).
A comparison of different models for estimating the height
in the corona from which such radio emission emanates shows
that the uncertainties in the height estimates can range from
5 to 10\% (Mann et al. 1999). 
The difference between the heights quoted in the third and fourth columns is relevant only for
the nonthermal radio emission at 200, 500 and 1060 MHz.

Masuda et al. (1994) estimate that the
hard X-ray emission at the top of the flaring loop originates at a height of 
around 18,000 km. We have used this number as a rough estimate of the height at which the hard 
X-ray flare (lower curve, figure 1) originates.  The fifth column gives the start time of
the phenomenon, as evident from the corresponding lightcurve.  We surmise that the nonthermal 
emission is a signature of magnetic reconnection, which provides sites for particle
acceleration and consequent nonthermal emission. The start of nonthermal emission at a particular frequency 
therefore signals the start of reconnection processes at the height corresponding to that
frequency. 

The first sign of nonthermal activity associated with this event is the 200 MHz nonthermal emission quoted in
item A (table 1). Detailed measurements of the trajectory of the prominence eruption associated with this event 
made by Kundu et al. (2003) using 17 GHz data from the NoRH reveal that the prominence is at $\sim$ 1.07 R$_{\odot}$ at 04:22 UT.
The 200 MHz nonthermal emission peak of item A (table 1) occurs at 1.08 (s=1) or 1.14 R$_{\odot}$ (s=2), which is
clearly above the prominence.
This could be consistent with the predictions of the breakout
model for solar eruptions (Antiochos et al. 1999), in which the first signature of eruption is the
null point reconnection above the central arcade which is going to erupt. According to this model, 
the central arcade rises slowly during the initial
buildup stage, due to
shearing at its footpoints, pushing the overlying constraining field lines
and distorting the X-type null point
(above the central arcade). Beyond a certain point, the current layer at the overlying null point becomes
sufficiently thin so that ideal MHD evolution is no longer possible and reconnection sets in, so that 
the overlying constraining field lines connect to adjoining arcades.  
The measurements of Kundu et al. (2003)
however reveal that the prominence had started rising well before 04:22 UT, the time at which the 
200 MHz nonthermal
emission peak of item A (table 1) occurs. This could be because the evolution of the null point above
the central arcade prior to 04:22 UT was ideal, since the current layer was not thin enough for reconnection
to set in. However, this does remain an open question. As mentioned earlier, it should also be borne in mind that there could be an uncertainty of the order of 5-10\% in the estimate of the height from which the 200 MHz emission originates. The uncertainty in the projected position of the filament in the 17 GHz images used by Kundu et al. (2003) is much smaller; it is of the order of 1 beam, which is around 15 arcseconds or 0.01 R$_{\odot}$. 

The next noteworthy feature evident from Table 1 is the 
near-simultaneity of items 
B, C, D, E, F and G. The 200, 410, 500 and 1060 MHz lightcurves exhibit peaks between $\sim$ 04:48 -- 04:50 UT.
The X-ray flare and the prominence eruption also take place in this time interval. 
Measurements of the prominence 
trajectory by Kundu et al. (2003) using NoRH 17 GHz data show that items B--G (Table 1) occur well below the prominence. The 
near-simultaneous nonthermal emission at heights ranging from 1.02 to 1.18 R$_{\odot}$ could thus arise 
from a vertical current sheet below the erupting arcade.  
During this time interval, 
the lower envelope of the drifting continuum in the HiRAS dynamic spectrum (see figures 4 and 5) could represent 
the top of this vertical current sheet. 
The velocity of this feature increases abruptly at around 
04:48 UT (right panel of figure 5), as is evident from the increase in the curvature of the lower
envelope of the drifting continuum around 04:48 UT (figure 4). This could be taken to mean that
the top of the vertical current sheet accelerates upward around this time. 
Our observations thus indicate that, during the interval 04:48 -- 04:50 UT:  
\begin{enumerate}
\item a vertical current sheet spanning at least 1.02 to 1.18 R$_{\odot}$
exists (items B--G, table 1)\\
\item the top of the current sheet accelerates upward (figure 5, right panel)\\
\item the prominence eruption and X-ray flare occur \\ 
\end{enumerate}
The formation of a vertical current sheet under the erupting arcade is a 
feature of several flux rope models for CME initiation (e.g., Forbes 1991; Lin \& Forbes 2000; 
Mikic \& Linker 1994; Amari et al. 2000). Reconnection at the vertical current sheet allows the eruption
to proceed in these flux rope models. The height of the flux rope undergoes a sudden jump as a result of 
the (equally sudden) formation of a current sheet underneath it (Lin et al. 1998; Forbes 2000). 
The formation of a vertical current sheet underneath the erupting arcade is also a feature of the breakout model, 
although it is a secondary process not integral to the breakout per se (Antiochos et al. 1999; Klimchuk 2000). 
In the breakout model, reconnection in the vertical current sheet results in an increase in the speed 
of the eruption. The strong heating associated with this current sheet reconnection is manifested as the X-ray flare.

Items H, I and J (table 1) are post-eruptive phenomena. 
In particular, the 1060 MHz emission from the NE-SW oriented
GMRT source (item J) is most likely caused by flare accelerated
electrons trapped in the NE-SW oriented loop system. This is manifested as type IV-like decimetric continuum
 emission on the Hiraiso dynamic spectrum.

\section{Conclusions}
We have presented some of the first GMRT observations of a solar eruption. We have used GMRT observations at 
1060 MHz in conjunction with data from the NoRH, the Hiraiso solar observatory and LASCO to investigate the
onset mechanism of a flare-CME event on Nov 17 2001. We find evidence for reconnection above the erupting
arcade that could be construed as evidence in favor of the magnetic breakout model.
We find evidence for the formation of a vertical current
sheet below the erupting prominence. An abrupt increase in the velocity of the top of the vertical 
current sheet is acompanied by the M2.8 X-ray flare and
the prominence eruption. The features associated with the formation of the vertical current sheet 
are consistent with the predictions of several variants of flux 
rope CME initiation models, as well as the magnetic breakout model.

\acknowledgements
We thank the staff of the GMRT that made these observations possible. 
GMRT is run by the National Centre for Radio Astrophysics of the Tata 
Institute of Fundamental Research, India. We gratefully acknowledge the open data use policy of the Nobeyama Solar
Radio Observatory and the Learmonth Radio Observatory and the Mount Wilson Observatory. 
SOHO is a project of international cooperation between
NASA and ESA. This research has made use of the CME catalog generated 
and maintained by NASA and The Catholic University of America in cooperation with the
Naval Research Laboratory. 
PS thanks Dr. Kuniko Hori for kindly supplying detailed data from the Hiraiso spectrograph and Dr. James Klimchuk for
several useful discussions. We thank the anonymous referee for several critical comments on an earlier version
of this manuscript that helped us substantially improve the interpretation of the observations.
The work of MRK, SMW and VIG was supported by NASA grants NAG5-11872, NAG5-12860 and NSF grant ATM 9909809.

\end{article}

\begin{thebibliography}{}
\bibitem{}Amari, T., Luciani, J. F., Mikic, Z., Linker, J. 2000, ApJ, 529, L49
\bibitem{}Ananthakrishnan, S and Pramesh Rao, A., 2002, Giant Metrewave Radio 
Telescope, in Proc. Int'l Conf. on Multicolour Universe, Eds. R.K.Manchanda
and B.Paul, TIFR, Mumbai, 233 pp. Available at 
http://www.gmrt.ncra.tifr.res.in/gmrt$_{-}$hpage/Users/doc/doc.html

\bibitem{}Antiochos, S. K., Devore, C. R. and Klimchuk, J. A. 1999, ApJ, 510, 485
\bibitem{}Aschwanden, M. J. \& Benz, A. O. 1995, ApJ, 438, 997
\bibitem{}Forbes, T. G. 1991, Geophys. Astrophys. Fluid Dynamics, 62, 15
\bibitem{}Forbes, T. G. 2000, JGR, 105, 23153
\bibitem{}Klimchuk, J. A. 2001, in ``Space Weather''
   (Geophysical Monograph 125), ed. P. Song, H. Singer, \& G. Siscoe (Washington:  Am. Geophys. Un.), 143
\bibitem{} Kundu, M. R., et al. 2003, submitted to the Astrophysical Journal
\bibitem{}Lin, J., Forbes, T. G., Isenberg, P. A., Demoulin, P. 1998, ApJ, 504, 1006
\bibitem{}Lin, J. \& Forbes, T. G. 2000, JGR, 105, 2375
\bibitem{}Mann, G., Jansen, F., MacDowall, R. J., Kaiser, M. L., Stone, R. G. 1999, A\&A, 348, 614
\bibitem{}Masuda, S., Kosugi, T., Hara, H., TSuneta, S., and Ogawara, Y. 1994, Nature, 371, 495
\bibitem{}Mikic, Z., \& Linker, J. 1994, ApJ, 430, 898
\end{thebibliography}
\end{document}